\documentclass{article}
\usepackage{spconf}
\usepackage{lipsum}
\usepackage{subcaption}
\usepackage{graphicx}
\usepackage{amsmath}
\usepackage{wrapfig}
\usepackage{amssymb}
\usepackage{cite}
\usepackage{xcolor}

\title{
Coarse-to-Fine: Progressive Image Compression for Semantically Hierarchical Classification
}

\name{
    Jungwoo Kim$^{1,2}$, 
    Jun-Hyuk Kim$^{3\dagger}$, 
    Jong-Seok Lee$^{1,2\dagger}$\thanks{$^\dagger$Corresponding authors.}
}
\address{
    $^{1}$School of Integrated Technology, Yonsei University \\ 
    $^{2}$BK21 Graduate Program in Intelligent Semiconductor Technology, Yonsei University \\
    $^{3}$Department of Artificial Intelligence, Chung-Ang University
}

\begin{document}

\maketitle

\begin{abstract}
    Recent advances in learned image compression (LIC) have enabled practical deployments, spurring active research into image compression for machines and progressive coding schemes. However, their integration remains under-explored: prior works on progressive machine codec predominantly target sample-level difficulty adaptation (\textit{i.e.}, easy-to-hard), without considering semantic-level scalability. In this work, we introduce a semantic hierarchy-aware progressive codec that enables semantic scalability (\textit{i.e.}, coarse-to-fine) from a single bitstream. We first systematically categorize ImageNet-1K classes into CLIP embedding-based semantic hierarchies. Based on a channel-wise autoregressive framework, we decompose latent representations into hierarchically ordered channel blocks, each explicitly optimized for a corresponding semantic hierarchy. Extensive experiments demonstrate that our approach substantially improves coarse-level recognition at low bitrates while maintaining fine-grained accuracy at higher bitrates. By reframing progressive transmission through the lens of semantic scalability, our work provides an efficient and interpretable solution for task-adaptive image coding, outperforming existing progressive codecs under hierarchical evaluation.
\end{abstract}

\section{Introduction}
\label{sec:intro}

Image compression, one of the fundamental research problems in image processing, has recently evolved from traditional signal processing techniques~\cite{skodras2001jpeg2000, bross2021overview} to learned image compression (LIC)~\cite{balle2018variational, lu2022transformer, kim2022joint, liu2023learned}, achieving superior rate-distortion performance through end-to-end optimization.
However, most LIC codecs lack fine-grained scalability, requiring re-encoding or separate models to operate at different bitrates. 
To address this limitation, progressive LIC~\cite{lu2021progressive, lee2022dpict, jeon2023context, hojjat2023progdtd, hojjat2024limitnet, presta2025efficient, kim2025progressive, lee2025deephq} has been studied recently, enabling incremental reconstruction from a single bitstream under bandwidth-constrained or latency-sensitive environments. 
While this progressive coding is particularly critical for image compression for machines (ICM)---where vision systems are typically deployed on autonomous platforms and edge devices---progressive coding for machine vision still remains under-explored.

\begin{figure}[t]
  \centering
  \begin{subfigure}[b]{0.485\linewidth}
    \centering
    \includegraphics[width=\linewidth]{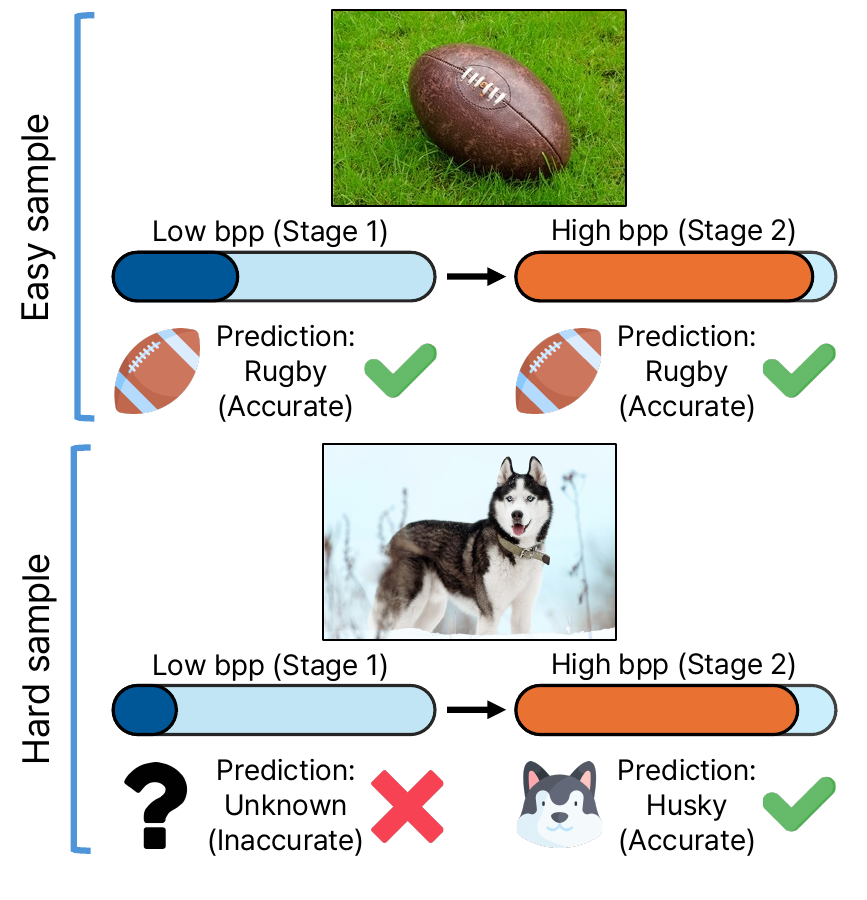}
    \caption{Easy-to-hard~\cite{hojjat2024limitnet, kim2025progressive}}
    \label{fig:example-1}
  \end{subfigure}
  \hfill
  \begin{subfigure}[b]{0.485\linewidth}
    \centering
    \includegraphics[width=\linewidth]{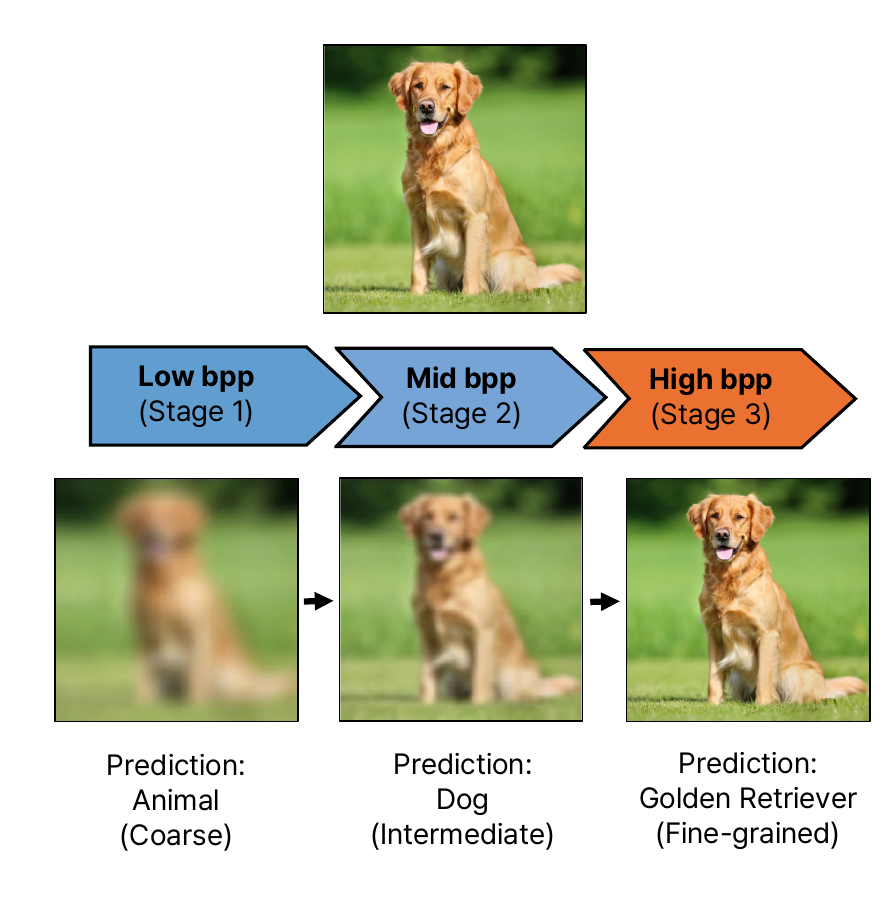}
    \caption{Coarse-to-fine (Ours)}
    \label{fig:example-2}
  \end{subfigure}
  \vspace{-0.5em}
  \caption{Conceptual comparison of the scalability in progressive image compression for machines.}
  \label{fig:example}
  \vspace{-1.0em}
\end{figure}

Existing progressive codecs are primarily designed for human perception, where degraded low-quality reconstruction is often sufficient for visual understanding. 
Machines, however, are highly sensitive to noises, thus performance degrades substantially at low bitrates, rendering early-stage outputs unreliable for downstream tasks. 
Existing works on progressive compression for machines~\cite{hojjat2024limitnet, kim2025progressive} have attempted to address this by adapting bitrate allocation to sample-level difficulty (\textit{i.e.}, easy-to-hard), transmitting harder samples with more bits (see Fig.~\ref{fig:example-1}).  
While this strategy improves average performance, it does not consider what information should be prioritized at each decoding stage, leading to severe accuracy drops at low bitrates.
To this end, we argue that aligning the progressive bitstream with a semantic hierarchy (\textit{i.e.}, coarse-to-fine) is a natural and more effective direction.

In this work, we propose a semantic hierarchy-aware progressive codec that aligns each decoding stage with a structured semantic hierarchy~\cite{park2025visually}. 
Our codec first transmits coarse, high-level semantics under tight bitrate constraints, and gradually injects fine-grained details with additional bits.
Within a channel-wise autoregressive framework~\cite{liu2023learned}, latent representations are decomposed into ordered channel blocks, each explicitly tied to a semantic level, from coarse taxonomy to fine-grained classes. 
Extensive experiments show that our method achieves strong coarse-level recognition under low bitrates and maintains competitive fine-grained accuracy at higher bitrates, outperforming existing state-of-the-art progressive codecs and demonstrating that semantic hierarchy-aware design is a promising direction for practical task-adaptive image coding.
\section{Related Work}
\label{sec:rework}

\noindent \textbf{Progressive Image Compression.}
Progressive image compression enables a single bitstream to be decoded at multiple levels for flexible rate control.
Early efforts primarily focused on training schemes to achieve scalability~\cite{lu2021progressive, hojjat2023progdtd, hojjat2024limitnet, lee2025deephq}, while more recent works introduced trit-plane coding~\cite{lee2022dpict, jeon2023context, kim2025progressive} and efficient latent ordering~\cite{presta2025efficient}.
Despite these technical advancements, they remain primarily optimized for visual fidelity, with limited consideration of task-level semantics.
While recent studies have emerged regarding progressive ICM~\cite{hojjat2024limitnet, kim2025progressive}, they primarily focused on adapting bitrate to sample-level difficulty, while still overlooking hierarchical semantic structures.

\vspace{0.5em}
\noindent \textbf{Image Compression for Machine.}
The landscape of image compression has shifted significantly from conventional standards to LIC, leveraging deep neural networks to achieve superior rate-distortion performance~\cite{balle2018variational, kim2022joint, liu2023learned}.
As LIC matures beyond human perceptual quality, research focus has increasingly expanded from human-centric reconstruction to ICM, where the goal is to preserve task-essential semantics.
Early ICM frameworks~\cite{le2021image, shindo2024image} typically relied on end-to-end joint optimization with task objectives, while recent studies have introduced task-adaptation modules~\cite{chen2023transtic, li2024image, zhang2024all}. 
Furthermore, the field is currently exploring the integration of generative models~\cite{lee2025slim} to synthesize task-relevant details from ultra-low bitrates. 

\section{Semantic Hierarchy in ImageNet}
\label{sec:hierarchy}

\begin{table}[t]
\centering
\caption{Quantitative evaluation of semantic coherence via Wu-Palmer (WUP) Similarity ($K=10$).}
\label{tab:wup}
\vspace{-0.5em}
\resizebox{\linewidth}{!}{
\begin{tabular}{l|ccc}
\hline
\textbf{Method} &  \textbf{$\text{WUP}_{\text{Intra}}$} ($\uparrow$) & \textbf{$\text{WUP}_{\text{Inter}}$} ($\downarrow$) & \textbf{Gap} \\ \hline
WordNet Depth-cut & 0.472 & 0.270 & $+$0.202 \\
CLIP Clustering & 0.583 & 0.426 & $+$0.157 \\ \hline
\end{tabular}
}
\vspace{-1.0em}
\end{table}

To enable semantically progressive transmission, we first establish a three-level hierarchy over the ImageNet-1K~\cite{ILSVRC15} classes.

\begin{figure}[t]
  \centering
  \includegraphics[width=\linewidth]{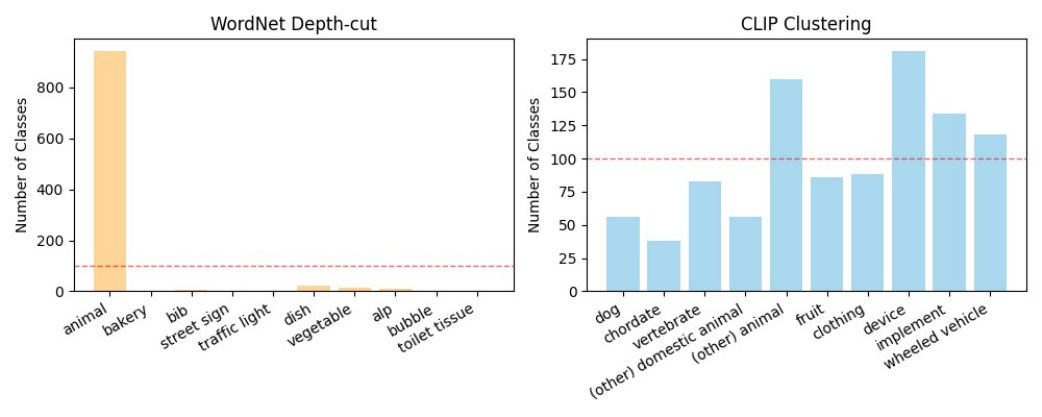}
  \vspace{-2.5em}
  \caption{Comparison of cluster size distributions for the coarse-level hierarchy. The WordNet Depth-cut (left) exhibits severe skewness, while our CLIP-based clustering (right) yields a balanced distribution.}
  \label{fig:class-histogram}
  \vspace{-1.5em}
\end{figure}

\vspace{0.5em}
\noindent \textbf{CLIP-based Semantic Clustering.}
Since ImageNet-1K classes are intrinsically mapped to WordNet~\cite{wordnet1994miller} synsets, a straightforward approach to establish a semantic hierarchy is to perform a depth-based cut on the WordNet tree, targeting a predefined number of clusters $K \in \{10, 100, 1000\}$, which correspond to the coarse, intermediate, and fine levels, respectively.
However, the inherent linguistic structure of WordNet does not translate to a balanced visual hierarchy.
As illustrated in Fig.~\ref{fig:class-histogram}, a depth-based cut (left) leads to a severe structural collapse; at a coarse granularity ($K=10$), most classes are subsumed under a single cluster (``animal").
Such extreme skewness renders a hierarchy ineffective for progressive transmission, as the coarse-level signal provides negligible discriminatory information. 

To overcome this imbalance, we propose a data-driven hierarchy derived from the embedding space of CLIP~\cite{radford2021learning}.
For each class, we generate a text embedding using the template ``a photo of $\{\;\}$".
We then apply k-means clustering to the normalized embeddings to partition them into clusters.
As shown in Fig.~\ref{fig:class-histogram}, our method produces a substantially more uniform distribution, ensuring that each semantic node carries sufficient information.

\begin{figure*}[t]
    \centering
    \begin{subfigure}{0.65\textwidth}
        \centering
        \includegraphics[width=\linewidth]{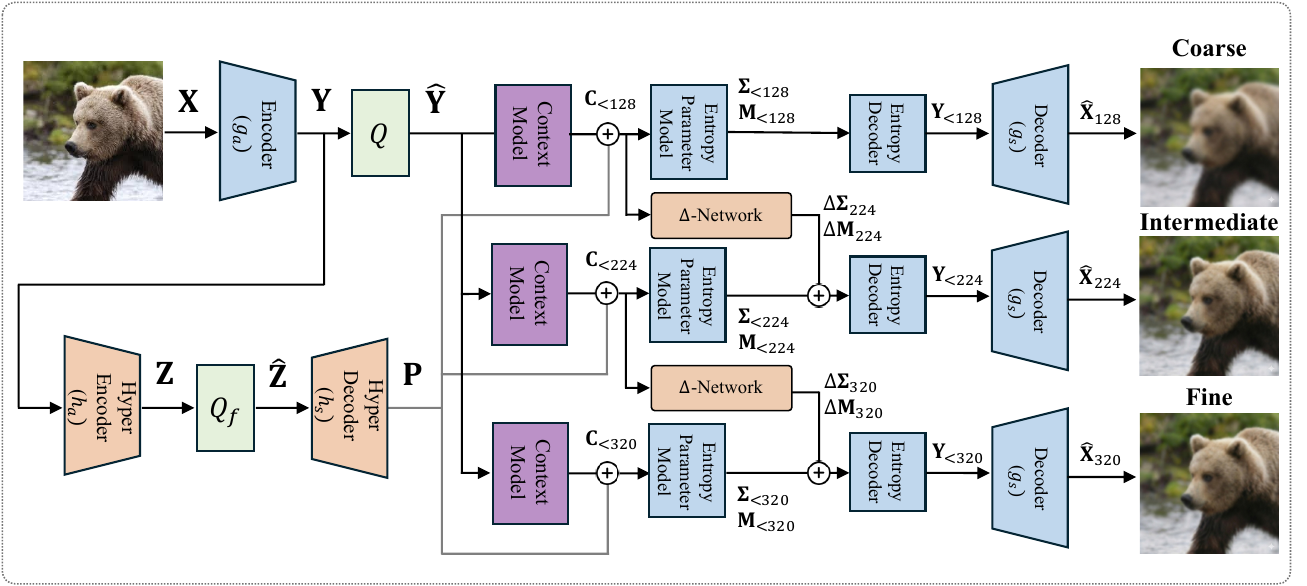}
        \caption{Overall framework}
        \label{fig:overall}
    \end{subfigure}
    \begin{subfigure}{0.3\textwidth}
        \centering 
        \includegraphics[width=\linewidth]{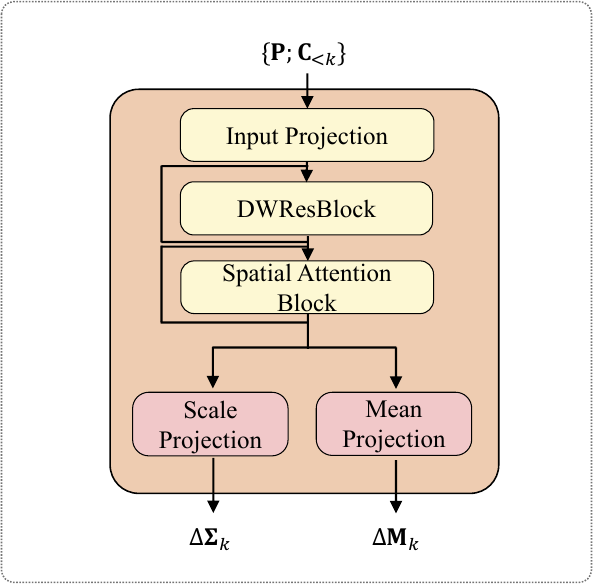}
        \caption{$\Delta$-Network}
        \label{fig:ppn}
    \end{subfigure}
    \vspace{-0.5em}
    \caption{Detailed pipeline of our proposed codec. Note that context model, entropy parameter model, entropy decoder, and decoder are shared across decoding level ($k)$.}
    \label{fig:framwork}
    \vspace{-1.0em}
\end{figure*}

\vspace{0.5em}
\noindent \textbf{Quantitative Validation via WUP Similarity.}
The semantic coherence of a cluster $G$ is quantified using the Wu-Palmer (WUP)~\cite{wu1994verb} similarity between classes ($c_i$):
\begin{equation}
    S_{\text{WUP}}(c_1,c_2)=\frac{2\cdot \text{depth}(\text{LCA}(c_1,c_2))}{\text{depth}(c_1)+\text{depth}(c_2)},
    \label{eq:wup}
\end{equation}
where $\text{LCA}(\cdot, \cdot)$ denotes the lowest common ancestor in the WordNet tree.
To verify if these clusters maintain semantic coherence, we evaluate the Intra-cluster and Inter-cluster WUP similarities.
As summarized in Table~\ref{tab:wup}, while the WordNet Depth-cut yields a higher gap (since it is optimized for the metric), our CLIP-based clustering demonstrates a robust alignment with visual semantics, achieving an Intra-WUP of 0.583 against an Inter-WUP of 0.426 at $K=10$.
Without any explicit taxonomic supervision, we effectively segregate fundamentally different categories such as ``Vertebrates" and ``Chordate", reflecting a meaningful capture of visual-semantic hierarchies.

\section{Methods}
\label{sec:methods}
In this section, we introduce our semantic hierarchy-aware progressive codec (see Fig.~\ref{fig:framwork}), which aligns each decoding stage with a corresponding level of semantic hierarchy.

\vspace{-0.8em}
\subsection{Systematic Overview}
\label{sec:overview}
Following the previous works~\cite{balle2018variational, cheng2020learned, lu2022transformer, kim2022joint, liu2023learned}, 
we adopt a hyperprior-based learned image compression framework.
An input image $\mathbf{X}$ is transformed into latent representation $\mathbf{Y}$ by an encoder $g_a$, and further compressed into a hyperlatent $\mathbf{Z}$ via a hyper encoder $h_a$.
Then, the quantized hyperlatent $\hat{\mathbf{Z}}$ is processed by a hyper decoder $h_s$ to produce hyperprior features $\mathbf{P}$.
A context model $\mathcal{C}$ captures spatial dependencies from already-decoded latents to generate context features $\mathbf{C}$.
The entropy parameter model $\mathcal{H}$ fuses $[\mathbf{P}; \mathbf{C}]$ to produce distribution parameters $[\mathbf{\Sigma},\mathbf{M}]$.
During reconstruction, decoder $g_s$ decodes the quantized latent $\hat{\mathbf{Y}}$ to obtain the reconstructed image $\hat{\mathbf{X}}$.
The bitrate is determined by the entropy of $\hat{\mathbf{Y}}$ and $\hat{\mathbf{Z}}$ under the learned probability models.

\vspace{0.5em}
\noindent \textbf{Semantic Alignment.}
In our codec, the latent representation $\mathbf{Y}$ consists of $C=320$ channels is grouped into three fixed prefix blocks: coarse, intermediate, and fine.
Formally, let $\hat{\mathbf{Y}}_{<k}$ be the prefix block of the first $k$ channels.
We set $k\in \{128, 224, 320\}$ for the coarse, intermediate, and fine semantic levels, respectively.
By decoding $\hat{\mathbf{Y}}_{<k}$ yields the reconstructed image $\hat{\mathbf{X}}_{<k}$ at the corresponding semantic level.
All stages share a single decoder $g_s$.

\vspace{0.5em}
\noindent \textbf{Latent Ordering.}
Within each prefix block, latent symbols are further organized according to the scale parameters predicted by the hyperprior~\cite{lu2021progressive, presta2025efficient, kim2025progressive}.
Specifically, each symbol---one coefficient at location $(c,h,w)$---is ordered by the magnitude of its predicted standard deviation $\sigma$ in $\mathbf{\Sigma}$.
Symbols with larger uncertainty are transmitted earlier, while those with smaller predicted scales are deferred to later positions.
As a result, each prefix not only corresponds to a semantic level but also captures the informative latent dimensions under the learned entropy model, yielding a rate-efficient progressive representation.

\vspace{0.5em}
\noindent \textbf{Training Objective.}
The codec is trained end-to-end with a combination of rate-distortion and task-specific losses for each semantic level $k$:
\begin{equation}
    \mathcal{L} = \sum_{k} \left[ \text{BPP}_k + \lambda_k \cdot \text{MSE}(\hat{\mathbf{X}}_{<k}, \mathbf{X}) + \gamma_k \cdot \mathcal{L}_{\text{CE}}^{(k)}(\hat{\mathbf{X}}_{<k}) \right],
\end{equation}
where $\mathcal{L}_{\text{CE}}^{(k)}$ is the cross-entropy loss evaluated at the semantic granularity corresponding to prefix $k$.
The MSE term is included to stabilize the training.
The parameters $\lambda_k$ and $\gamma_k$ are hyperparameters used to control the balance between different loss terms during training.

\begin{figure*}[t]
  \centering
  \begin{subfigure}{0.25\textwidth}
      \centering
      \includegraphics[width=\linewidth]{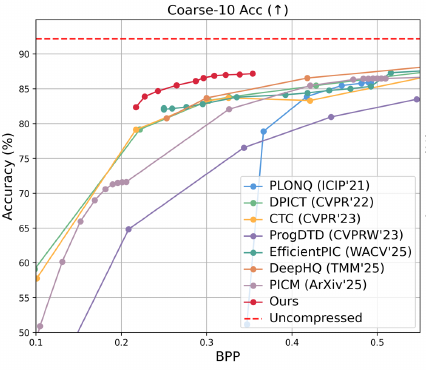}
      \vspace{-1.7em}
      \caption{Coarse-level accuracy}
      \label{fig:hier-acc-coarse}
  \end{subfigure}
  \begin{subfigure}{0.25\textwidth}
      \centering 
      \includegraphics[width=\linewidth]{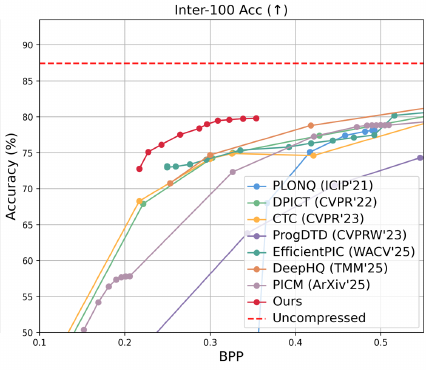}
      \vspace{-1.7em}
      \caption{Inter-level accuracy}
      \label{fig:hier-acc-inter}
  \end{subfigure}
  \begin{subfigure}{0.25\textwidth}
    \centering 
    \includegraphics[width=\linewidth]{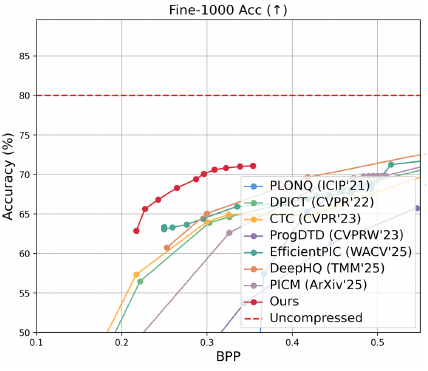}
    \vspace{-1.7em}
    \caption{Fine-level accuracy} 
    \label{fig:hier-acc-fine}
\end{subfigure}
  \vspace{-0.5em}
  \caption{Hierarchical classification performance in each semantic level with ResNet-50~\cite{he2016deep}.}
  \label{fig:accuracy}
  \vspace{-1.0em}
\end{figure*}

\begin{figure*}
\centering
\begin{subfigure}{0.25\textwidth}
    \centering
    \includegraphics[width=\linewidth]{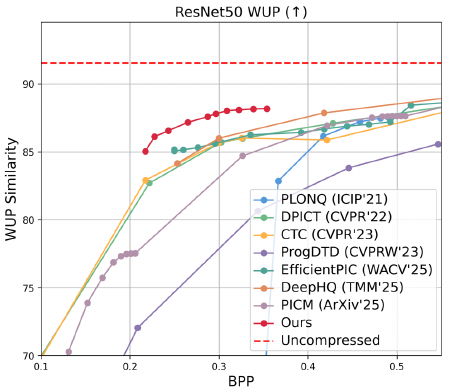}
    \vspace{-1.7em}
    \caption{ResNet-50~\cite{he2016deep}}
    \label{fig:resnet-wup}
\end{subfigure}
\begin{subfigure}{0.25\textwidth}
    \centering 
    \includegraphics[width=\linewidth]{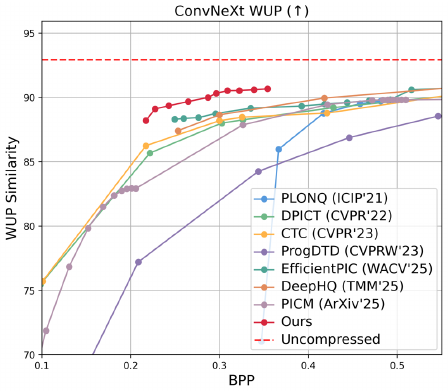}
    \vspace{-1.7em}
    \caption{ConvNeXt-Tiny~\cite{liu2022convnext}}
    \label{fig:convnext-wup}
\end{subfigure}
\begin{subfigure}{0.25\textwidth}
  \centering 
  \includegraphics[width=\linewidth]{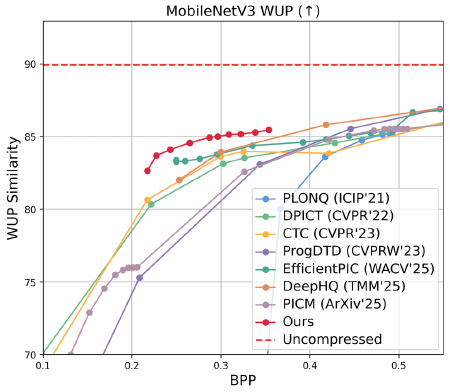}
  \vspace{-1.7em}
  \caption{MobileNetV3~\cite{howard2019mobile}} 
  \label{fig:efficient-wup}
\end{subfigure}
\vspace{-0.5em}
\caption{Comparison of WUP Similarity~\cite{wu1994verb} with different downstream task models.}
\label{fig:wup-comparison}
\vspace{-1.5em}
\end{figure*}

\vspace{-0.8em}
\subsection{Level-wise Autoregressive Refinement}
\label{sec:ar-refine}
While the fixed channel ordering aligns latent prefixes with semantic levels, the hyperprior predicts one set of parameters for all channels.
To better capture the semantic scalability, we introduce a level-wise autoregressive refinement module, which adaptively adjusts the entropy parameters for each semantic block based on the information from preceding blocks.

\vspace{0.5em}
\noindent \textbf{Prefix-Masked Context Modeling.}
For each prefix level $k \in \{128, 224, 320\}$, we construct context features that reflect only the already-transmitted channels.
Specifically, we form a prefix-masked latent representation:
\begin{equation}
    \tilde{\mathbf{Y}}_{<k} = \mathbf{I}_k \odot \hat{\mathbf{Y}},
\end{equation}
where $\mathbf{I}_k$ is a binary mask that zeros out channels beyond $k$.
Applying the context model $\mathcal{C}$ to $\tilde{\mathbf{Y}}_{<k}$ yields the prefix-aware context features:
\begin{equation}
    \mathbf{C}_{<k} = \mathcal{C}(\tilde{\mathbf{Y}}_{<k}).
\end{equation}
The base entropy parameters are then obtained by combining the hyperprior features $\mathbf{P}$ (derived from $\hat{\mathbf{Z}}$) with $\mathbf{C}_{<k}$:
\begin{equation}
    [\mathbf{\Sigma}_{<k}; \mathbf{M}_{<k}] = \mathcal{H}([\mathbf{P}; \mathbf{C}_{<k}]),
\end{equation}
where $\mathcal{H}$ denotes the entropy parameter network.

\vspace{0.5em}
\noindent \textbf{$\mathbf{\Delta}$-Networks.}
To enable adaptive refinement at semantic transitions, we introduce lightweight $\Delta$-networks $f_k$ for $k\in \{224,320\}$, which predict additive corrections to the base distribution parameters.
Each $f_k$ takes the features $[\mathbf{P};\mathbf{C}_{<k}]$ and learns channel-wise adjustments through depthwise residual blocks and spatial attention block~\cite{dosovitskiy2021an}.
These blocks efficiently capture spatial dependencies and channel interactions, enabling the network to adaptively modulate the distribution parameters based on semantic information from preceding channels:
\vspace{-0.5em}
\begin{equation}
    [\Delta \mathbf{\Sigma}_{k}; \Delta\mathbf{M}_{k}] = f_k([\mathbf{P};\mathbf{C}_{<k}]).
\end{equation}
\vspace{-0.5em}
The refined parameters are obtained via residual updates:
\begin{equation}
    \mathbf{\Sigma}_{<k}' = \mathbf{\Sigma}_{<k} + \Delta \mathbf{\Sigma}_{k}, \quad \mathbf{M}_{<k}' = \mathbf{M}_{<k} + \Delta \mathbf{M}_{k},
\end{equation}
where $\mathbf{\Sigma}_{<k}'$ is constrained to remain positive definite.
These refined parameters $[\mathbf{\Sigma}_{<k}', \mathbf{M}_{<k}']$ define the Gaussian distributions used for entropy coding and decoding of the corresponding channel prefix $\hat{\mathbf{Y}}_{<k}$, enabling more accurate rate estimation and efficient compression.

\vspace{-0.5em}
\section{Experiments}
\label{sec:experiments}

\subsection{Experimental Setup}
\noindent \textbf{Training Details.}
We train our codec on 80K images randomly sampled from the ImageNet-1K training set~\cite{ILSVRC15} for 100 epochs with a batch size of 8.
Our model is based on TIC~\cite{lu2022transformer}, with modifications to incorporate $\Delta$-networks that adaptively adjust distribution parameters at each semantic level.
We set $\mathbf{\lambda}=\{1e^{-4},1e^{-3},1e^{-2}\}$ and $\mathbf{\gamma}=\{0.5, 1.0, 2.0\}$ for rate-distortion trade-offs at each prefix level, and the cross-entropy loss is measured with ResNet-50~\cite{he2016deep}.
To prevent overly sharp gradients, we apply label smoothing in each prefix.

\noindent \textbf{Baselines.}
We compare our codec with existing state-of-the-art progressive LIC codecs: PLONQ~\cite{lu2021progressive}, DPICT~\cite{lee2022dpict}, CTC~\cite{jeon2023context}, ProgDTD~\cite{hojjat2023progdtd}, EfficientPIC~\cite{presta2025efficient}, DeepHQ~\cite{lee2025deephq}, and PICM~\cite{kim2025progressive}.
Note that progressive compression for machines remains under-explored; among these baselines, only PICM~\cite{kim2025progressive} is machine-oriented, while the rest are designed for human perception. 
Although LimitNet~\cite{hojjat2024limitnet} also targets progressive coding for machine vision, it focuses primarily on efficiency and latency rather than task performance, making meaningful rate-distortion comparisons difficult, we therefore exclude it from our baselines.

\begin{figure*}[t]
    \centering
    \includegraphics[width=0.9\linewidth]{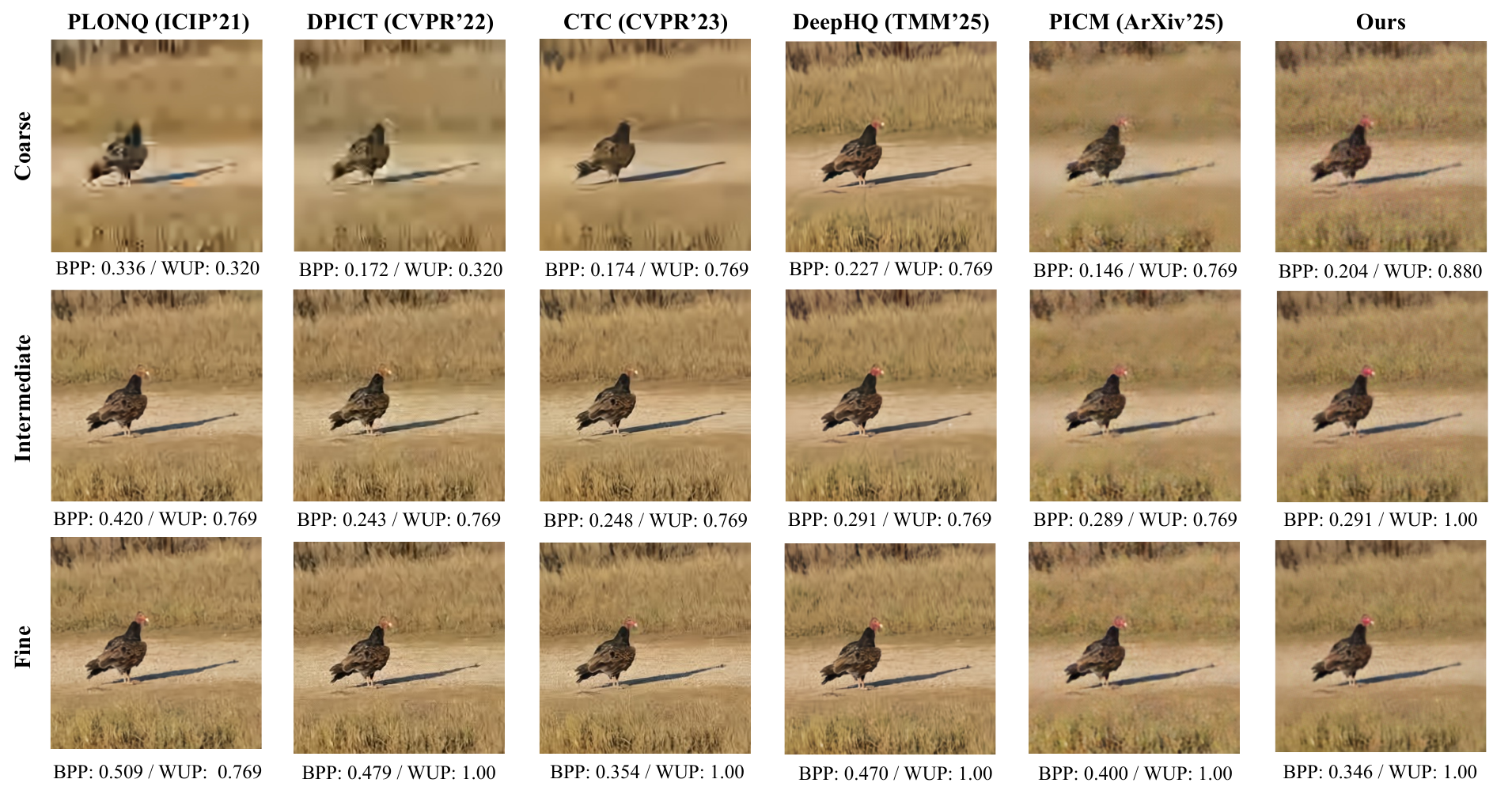}
    \vspace{-0.8em}
    \caption{Qualitative comparison of our codec against other baselines. The ground truth class is ``Vulture''. The Wu-Palmer similarity scores shown in the figure correspond to the following predicted classes: ``Airliner'' (WUP=0.320), ``Black grouse'' (WUP=0.769), and ``Kite (prey of bird)'' (WUP=0.880).}
    \label{fig:qualitative}
    \vspace{-1.5em}
\end{figure*}

\noindent \textbf{Evaluation.}
We evaluate hierarchical classification performance using ResNet-50~\cite{he2016deep} across coarse $(K=10)$, intermediate $(K=100)$, and fine $(K=1000)$ semantic levels.
Beyond accuracy, we also measure WUP similarity~\cite{wu1994verb} (see Eq.~\ref{eq:wup}), which quantifies how semantically close the predicted class is to the ground truth.
To demonstrate the robustness of our approach across different downstream task models, we evaluate WUP similarity with three different models: ResNet-50~\cite{he2016deep}, ConvNeXt-Tiny~\cite{liu2022convnext}, and MobileNetV3~\cite{howard2019mobile}.
All evaluations are conducted on the ImageNet-1K validation set.
All experiments are done with NVIDIA RTX 4090 GPUs.

\vspace{-0.5em}
\subsection{Results}
For our codec, we progressively decode the bitstream within the fixed prefix blocks: $C=[128, 320]$, where each symbol within each block is reordered according to its importance score.
As shown in Fig.~\ref{fig:accuracy}, our codec achieves superior performance at all three semantic levels---coarse, intermediate, and fine.
Notably, at the same low bitrate, our method outperforms baselines at the coarse level, demonstrating effective early semantic recovery.
Furthermore, as illustrated in Fig.~\ref{fig:resnet-wup}, our approach not only improves top-1 accuracy but also produces predictions that are semantically closer to the ground truth.

Beyond the ResNet-50 classifier used during training, we validate generalization across diverse classifier architectures.
Fig.~\ref{fig:convnext-wup} shows consistent improvements on the transformer-based ConvNeXt-Tiny~\cite{liu2022convnext}, and Fig.~\ref{fig:efficient-wup} also demonstrates gains on the lightweight MobileNetV3~\cite{howard2019mobile}, highlighting that progressive decoding is especially beneficial for these resource-constrained downstream machines.
This demonstrates that our learned semantic hierarchy transfers effectively across model families and computational budgets.
Also, as shown in Fig.~\ref{fig:qualitative}, our codec qualitatively produces more semantically coherent reconstructions compared to baselines.

\vspace{-0.5em}
\subsection{Ablation Study}
We ablate the design choises in our codec.
First, incorporating the $\Delta$-refinement network yields a BD-rate~\cite{bdrate} of $-1.401\%$ over the base codec, indicating meaningful bitrate savings at equivalent quality.

Also, to justify our channel allocation $\{128, 96, 96\}$, we compare against intermediate-heavy $\{96, 128, 96\}$, fine-heavy $\{96, 96, 128\}$, and uniform $\{107, 107, 106\}$ splits under the same total dimension of 320; the coarse-heavy split (ours) consistently achieves the best performance (ResNet-50: 59.25, WUP: 0.8352), confirming that allocating more channels to coarse semantics improves generalization across classifiers.

\section{Conclusion}
\label{sec:conclusion}
In this work, we presented a semantic hierarchy-aware progressive codec that aligns each decoding stage with a corresponding level of class granularity, enabling coarse-to-fine semantic scalability from a single bitstream. 
Experiments showed that reframing progressive transmission through semantic scalability outperforms existing codecs under hierarchical evaluation, offering a promising direction for task-adaptive image coding.

\vspace{0.5em}
\noindent \textbf{Future Works.}
While we demonstrate the effectiveness of semantic hierarchy-aware progressive coding, we employ only three-level hierarchy alignment with a relatively large latent dimension ($C=320$), which incurs overhead from the hyperprior $\mathbf{Z}$.
Exploring finer-grained semantic alignment and optimizing the bit allocation strategy to reduce this overhead present promising directions for future research.

\section{Acknowledgments}
This work was supported by the National Research Foundation of Korea (NRF) grant (No. RS-2024-00453301, RS-2025-00517159) and by Institute of Information \& communications Technology Planning \& Evaluation (IITP) grant (IITP-2025-RS-2024-00428780).

\newpage
\bibliographystyle{IEEEbib}
\bibliography{refs}

\end{document}